\let\TeXyear\year
\documentclass{ieeeaccess}

\let\year\TeXyear
\usepackage{cite}
\usepackage{graphicx}
\usepackage{textcomp}

\usepackage[most]{tcolorbox}
\usepackage{tikz} 
\newtcolorbox{highlighted}{colback=yellow,coltext=black,breakable}
\definecolor{accessblue}{cmyk}{1, 0.3, 0, 0.2}
\definecolor{greycolor}{cmyk}{0,0,0,.8}
\usepackage[T1]{fontenc}
\usepackage{lmodern}
\usepackage{braket}
\usepackage[version=3]{mhchem} %
\usepackage{here}
\usepackage{amsfonts,amsmath,amssymb,amsthm}
\usepackage{todonotes}
\usepackage{bm}
\usepackage{bbm}
\usepackage{autobreak}
\usepackage[noend]{algpseudocode}
\usepackage{xcolor}
\usepackage{authblk}

\usepackage{setspace}
\usepackage[font={sf,small,stretch=0.84},
            labelfont={bf,color=accessblue}
            ]{caption}
\usepackage{lipsum} 
\usepackage{url} 
\usepackage{multirow}

\DeclareMathOperator{\tr}{tr}

\usetikzlibrary{quantikz2}

\def\BibTeX{{\rm B\kern-.05em{\sc i\kern-.025em b}\kern-.08em
    T\kern-.1667em\lower.7ex\hbox{E}\kern-.125emX}}
\begin{document}

\title{SU(4) gate design via unitary process tomography: its application to cross-resonance based superconducting quantum devices}

\author{
\uppercase{Michihiko~Sugawara}\authorrefmark{1,2}, and
\uppercase{Takahiko~Satoh}\authorrefmark{3}
}
\address[1]{Quantum Computing Center, Keio University, Hiyoshi 3-14-1, Kohoku-ku, Yokohama, Kanagawa, 223-8522, Japan}
\address[2]{Graduate School of Science and Technology, Keio University, Hiyoshi 3-14-1, Kohoku, Yokohama 223-8522, Japan}
\address[3]{Department of Information and Computer Science, Keio University, Hiyoshi 3-14-1, Kohoku-ku, Yokohama 223-8522, Japan}

\corresp{Corresponding author: Michihiko Sugawara (email: sugawara.a6@keio.jp).}

\begin{abstract}
We present a novel approach for implementing pulse-efficient SU(4) gates on cross resonance (CR)-based superconducting quantum devices.
Our method introduces a parameterized unitary derived from the CR-Hamiltonian propagator, which accounts for static-$ZZ$ interactions. 
Leveraging the Weyl chamber's geometric structure, we successfully realize a continuous 
2-qubit basis gate, $R_{ZZ}(\theta)$, as an echo-free pulse schedule on the IBM Quantum device  {\tt ibm\_kawasaki}. 
We evaluate the average fidelity and gate time of various SU(4) gates generated using the $R_{ZZ}(\theta)$ to confirm the advantages of our implementation.
\end{abstract}


\titlepgskip=-15pt

\maketitle
\section{Introduction}\label{sec:Introduction}
Quantum computers have the potential to revolutionize various fields, including finance, quantum chemistry, optimization problems, and machine learning. 
In recent years, the competition to improve the performance of NISQ devices has made significant progress toward the realization of early fault-tolerant quantum computers~\cite{katabarwa2024early}, and quantum devices with more than 100 qubits have emerged in superconducting systems~\cite{kim2023evidence,ai2024quantum,gao2025establishing}. 
For instance, the coherence time of transmon-based superconducting qubits has now surpassed 100 $\mu$s, 
and the error rate of a single-qubit gate has reached a level of $10^{-4}$.
However, the error rate of 2-qubit gates remains at approximately $10^{-3}$, 
compared to that of single-qubit gates. 
Thus, improving the fidelity of 2-qubit gates has become an urgent priority for enhancing the capabilities 
of superconducting quantum devices.

\begin{figure*}[htbp]
\centering
\includegraphics[width=\textwidth]{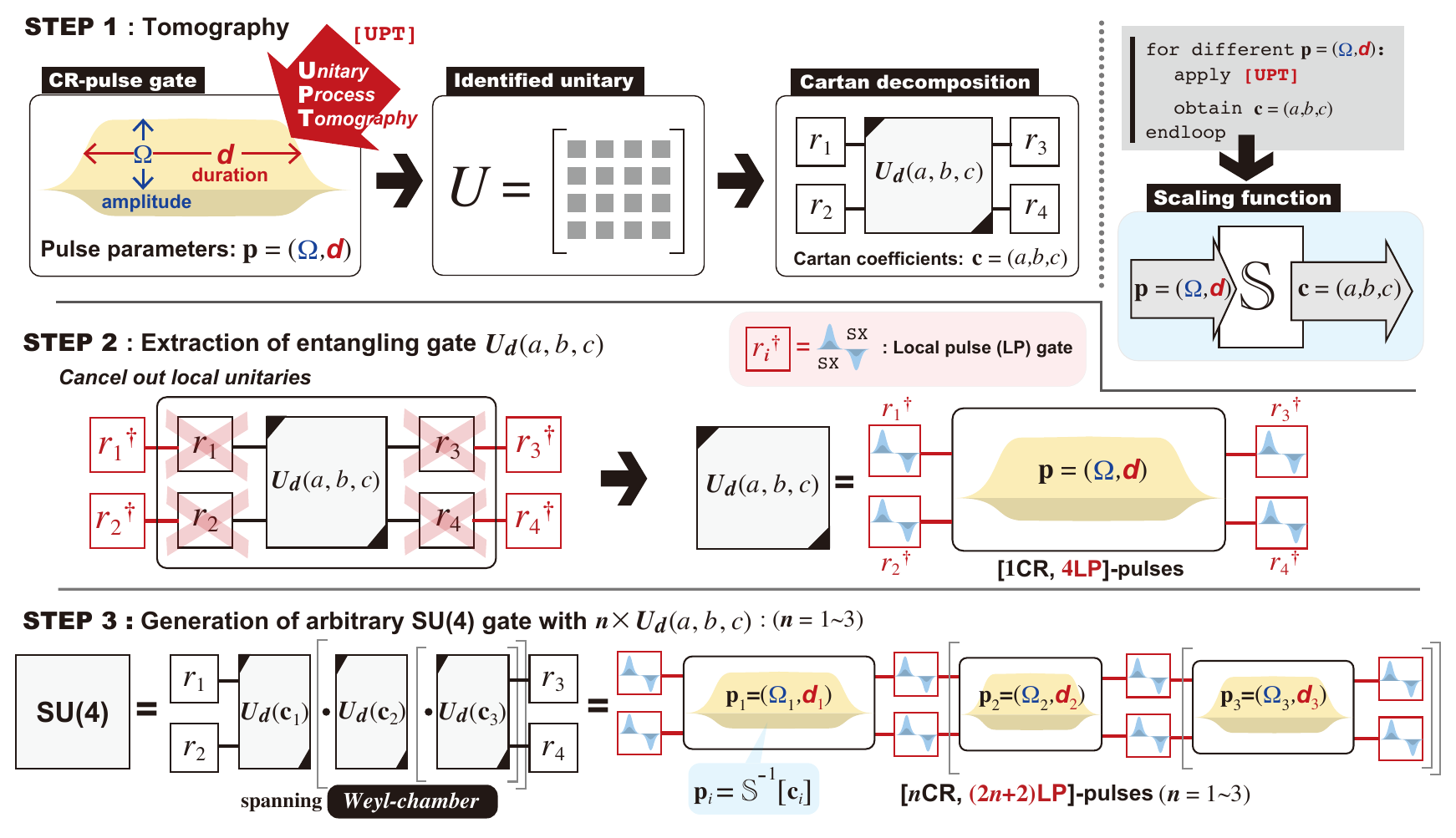}
\caption{Conceptual flow of the present work. 
$U_d(a,b,c)$ denotes the pure entangling gate defined in Eq.~(\ref{eq:Ud}) with Cartan coefficients $\mathbf{c} = (a,b,c)$ 
and $\{r_i\}~(i=1, \cdots, 4) \in SU(2)$ are local single-qubit operations.
}
\label{fig:concept}
\end{figure*}

Entangling gates on transmon qubits are generally implemented 
through electric dipole interactions.
One viable method employs the dynamics of a 2-qubit system driven by the effective device Hamiltonian, 
applicable under specific conditions
~\cite{rigetti2010fully, chow2011simple}.
Cross resonance (CR) scheme is one such condition, 
in which a microwave with the transition frequency of the control qubit is applied 
to the target qubit, while these qubits are connected through a weak static interaction
~\cite{sheldon2016procedure, magesan2020effective}.
This approach uses only microwave control, 
which allows us to use fixed frequency transmons, thereby reducing possible decoherence processes.
The CR scheme is widely employed in IBM Quantum devices, where the controlled-X (CX) or echoed CR (ECR) gate is provided as the sole 2-qubit basis gate~\cite{ecrgate}. 
Both CX and ECR gates are physically realized using the echoed CR pulse schedule~\cite{sheldon2016procedure, magesan2020effective, ecrgate}. 
While having a single, echo-based 2-qubit basis gate simplifies calibration process, it has certain drawbacks. 
For example, gate times are relatively long because the CR dynamics are inherently slow since it utilizes the weak static 
$ZZ$-interaction~\cite{rigetti2010fully, chow2011simple, magesan2020effective}. 
Moreover, some parameterized 2-qubit gates require decomposition into one to three CX (or ECR) gates, consuming valuable device coherence time.
To relieve such difficulties, we have shown that the total gate time for specific gates can be reduced 
by introducing a pulse-efficient ctrl-$\sqrt{\mathrm{X}}$ as an additional 2-qubit basis gate
~\cite{satoh2022pulse}. 
A more versatile solution is the implementation of a continuous basis gate that generates pulse-efficient 2-qubit gates based on their entangling capabilities.
Various types of such basis gates have been proposed
~\cite{stenger2021simulating, earnest2021pulse, chadwick2023efficient, selvan2023scheme, egger2023study, peterson2022optimal, niu2022pulse, niu2022effects}
with the help of pulse manipulation libraries supplied for the Qiskit~\cite{mckay2018qiskit, wille2019ibm, kanazawa2023qiskit}. 

Another drawback of the echoed pulse scheme is that the remaining static $ZZ$-interaction 
could be a potential source of unwanted systematic unitary errors. 
In the default echoed pulse implementation of CX(ECR), 
such unitary errors are ignored. 
Several attempts have been reported to resolve this issue
~\cite{sundaresan2020reducing, wei2021quantum}.
The most promising way is to apply the rotary echo tone pulses on target qubits which cancels the major part of unitary errors 
due to static $ZZ$-interaction and even brings the suppression of spectator errors
~\cite{sundaresan2020reducing}.
While the use of rotary tone pulses is a well-designed approach, the overall pulse schedule tends 
to become complicated due to the complex active control pulses on the target qubit. 
Additionally, the calibration process for these rotary tones becomes cumbersome.

In this study, to overcome the above difficulties, we aim to propose an echo-free continuous 2-qubit basis gate, $R_{ZZ}(\theta)$
on CR-based superconducting devices.
The essential concept behind this approach is depicted in Fig.~\ref{fig:concept}. 
We first apply the unitary process tomography (UPT) to identify the 2-qubit unitary matrix $U$ corresponding to a given crude CR-pulse gate, 
which consists of a single CR-pulse with pulse parameters, such as amplitude $\Omega$ and duration $d$.
Then, we apply the Cartan decomposition to the identified $U$ as
\begin{align}
    \label{eq:cartan_decomposition}
    U = (r_3 \otimes r_4) \cdot U_d(a,b,c) \cdot (r_1 \otimes r_2), 
\end{align}
with the Cartan coefficients $(a,b,c)$, 
where 
\begin{align}
    \label{eq:Ud}
    U_d(a, b, c) \equiv e^{-\frac{i}{2}(a XX + b YY + c ZZ)}, 
\end{align}
and $\{r_i\}~(i=1, \cdots, 4) \in SU(2)$ are local single-qubit operations. 
We omit tensor product sign $\otimes$ between two Pauli operators $\{X, Y, Z\}$ for simplicity.
Next, we investigate the relation between the Cartan coefficients $\mathbf{c}\equiv(a, b, c)$ and CR-pulse parameters $\mathbf{p} \equiv (d, \Omega)$, 
by scanning pulse parameters and applying UPT. 
The relation could be expressed as the scaling function as $\mathbf{c} = \mathbb{S}[\mathbf{p}]$ and its inverse function $\mathbf{p} = \mathbb{S}^{-1}[\mathbf{c}]$ as well.
As STEP 2, we apply local gates to cancel out the local unitary components to extract $U_d(a,b,c)$ gate.
Since each local gate $r_i$ is decomposed into 2 {\tt SX} basis gates, implemented with two $\pi/2$-pulses, 
$U_d(a,b,c)$ is given as a simple pulse sequence consisting of a single CR-pulse and 4 local pulses (8 {\tt SX}), 
which we abbreviate as [1CR, 4LP]. 
As STEP 3, we construct an arbitrary SU(4) gate with 1 to 3 $U_d(a,b,c)$ 
~\cite{PhysRevA.67.042313, earnest2021pulse, peterson2022optimal} together with the scaling function $\mathbb{S}^{-1}[\mathbf{c}]$, 
which is realized as $[n {\rm CR}, (2n+2){\rm LP}]$-pulses. 

The rest of this paper is organized as follows.
In Sec.~\ref{sec:Preliminaries}, we briefly introduce the prerequisites for this research, such as the Weyl chamber, 
native 2-qubit gate of the effective CR-Hamiltonian and the UPT.
In Sec.~\ref{sec:Results}, we show the experimental results on the IBM Quantum superconducting device, {\tt ibm\_kawasaki}, highlighting fidxtraclity and gate time improvements. 
Section~\ref{sec:Discussions} discusses the implications of our findings and potential applications and the paper is concluded with a summary and future outlook in Sec.~\ref{sec:Conclusion}. 

\section{Preliminaries}\label{sec:Preliminaries}
\subsection{Weyl chamber and $\boldsymbol{R_{ZZ}(\theta)}$ as a 2-qubit basis gate}

\begin{figure*}[ht]
\centering
\includegraphics[clip,width = 14cm]{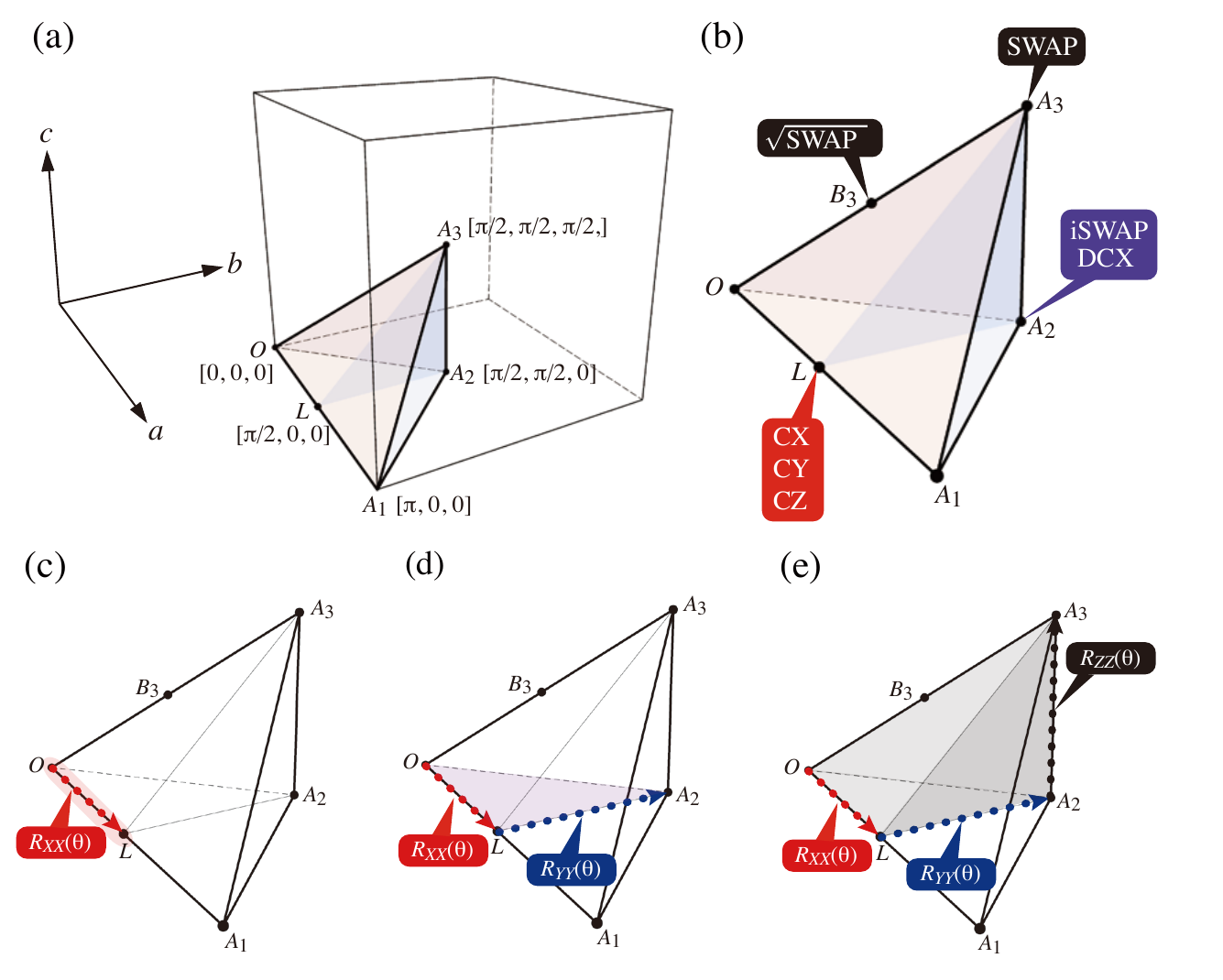}
\caption{(a) Weyl chamber (tetrahedron \(OA_1A_2A_3\)) contains all the locally equivalent class of 2-qubit operations.
(b) Four important points in the Weyl chamber; at each point, typical locally equivalent gates are indicated. 
DCX denotes the double-CX gate, which is a 2-qubit gate composed of two back-to-back CX gates with alternate controls. 
$B_3$ is defined as $B_3=[\pi/4,\pi/4,\pi/4]$.
(c) The line $OL$ contains all the locally equivalent gates that can be generated from the continuous basis gate $R_{XX}(\theta)$. 
(d) The triangle $OLA_2$ contains SU(4) gates that can be generated with $R_{XX}(\theta)$ and $R_{YY}(\theta)$.
(e) The gray area $OLA_2A_3$ contains SU(4) gates that can be generated with $R_{XX}(\theta)$, $R_{YY}(\theta)$ and $R_{ZZ}(\theta)$.
}
\label{fig:Weyl-Chamber}
\end{figure*}

The Cartan decomposition, Eq.~(\ref{eq:cartan_decomposition}), is closely related to the Weyl chamber, 
which provides a clear view of the geometric structure of 2-qubit gates
~\cite{PhysRevA.67.042313, watts2013metric, kraus2001optimal, musz2013unitary, zhang2004optimal, 
muller2011optimizing, goerz2017charting, zhang2005conditions, balakrishnan2010entangling, rezakhani2004characterization}. 
The Weyl chamber is defined as a tetrahedron \(OA_1A_2A_3\) in 
Fig.~\ref{fig:Weyl-Chamber}(a), which contains all the 
locally equivalent class of 2-qubit operations, represented by the point $[a, b, c]$. 
Here, we refer to 2-qubit unitary operators $U$ and $V$ as being locally equivalent 
when they satisfy $U = (r_3 \otimes r_4) \cdot V \cdot (r_1 \otimes r_2)$ with local operations $\{r_i\}\, (i=1,2,3,4)$.
Shown in Fig.~\ref{fig:Weyl-Chamber}(b) are particularly important points corresponding to familiar 2-qubit gates. 
Each point within the Weyl chamber is characterized by the entangling power (EP), 
which is the metric of mean entanglement generated by a corresponding 2-qubit gate  
~\cite{balakrishnan2010entangling, rezakhani2004characterization}.
From the Cartan coefficients $[a, b, c]$, one can readily calculate EP as
\begin{align}
\label{eq:EP}
    {\rm EP}&(a,b,c) = -\frac{1}{18} \cos(2a) \cos(2b) \nonumber \\
            &-\frac{1}{18} \cos(2c) \cos(2b) 
             -\frac{1}{18} \cos(2a) \cos(2c) + \frac{1}{6}. 
\end{align}
EP takes a maximum value 2/9 at $[0,0,\pi/2]$ and $[\pi/2, \pi/2,0]$, while a minimal value 0 at $[0,0,0]$ and $[\pi/2, \pi/2, \pi/2]$. 

The geometric structure of the Weyl chamber is useful in elucidating the concept of a 2-qubit basis gate for implementing universal SU(4) gates.
Since $XX$, $YY$ and $ZZ$ commute with each other, 
$U_d(a,b,c)$ is written as 
\begin{align}
    U_d(a,b,c) = R_{XX}(a) \cdot R_{YY}(b) \cdot R_{ZZ}(c),
\end{align}
where $R_{AB}(\theta) \equiv \exp[-i (\theta/2) A B]$ with $A,B \in \{X,Y,Z\}$. 
Thus, Weyl chamber can be spanned with $R_{XX}(a)$, $R_{YY}(b)$, and $R_{ZZ}(c)$. 
If we choose $R_{ZZ}(\theta)$ as a basis gate, $R_{XX}(\theta)$, $R_{YY}(\theta)$ are derived from $R_{ZZ}(\theta)$ as 
\begin{align}
    R_{XX}(\theta) &= \{r_Y(\mp \pi/2) \otimes r_Y(\mp \pi/2)\} \cdot \nonumber \\
    &R_{ZZ}(\theta) \cdot \{r_Y(\pm \pi/2) \otimes r_Y(\pm \pi/2)\}, \label{eq:XX_from_ZZ}\\
    R_{YY}(\theta) &= \{r_X(\mp \pi/2) \otimes r_X(\mp \pi/2)\} \cdot \nonumber \\
    &R_{ZZ}(\theta) \cdot \{r_X(\pm \pi/2) \otimes r_X(\pm \pi/2)\}, \label{eq:YY_from_ZZ}
\end{align}
where $r_A(\theta) \equiv \exp [-i (\theta/2) A]$.
Since $R_{ZZ}(\theta)$ belongs to locally equivalent class $[\theta, 0, 0]$, 
a single continuous basis gate $R_{ZZ}(\theta)$ can generate SU(4) gates on the $a$-axis in the Weyl chamber 
including arbitrary ctrl-$U$ gates as shown in Fig.~\ref{fig:Weyl-Chamber}(c). 
Similarly, we can create locally equivalent gates belonging to the plane $OLA_2$ in Fig.~\ref{fig:Weyl-Chamber}(d) with 2 $R_{ZZ}(\theta)$ gates
using Eqs.~(\ref{eq:XX_from_ZZ}) and (\ref{eq:YY_from_ZZ}), 
and the gates in the gray area $OLA_2A_3$ shown in Fig.~\ref{fig:Weyl-Chamber}(e) with 3 $R_{ZZ}(\theta)$. 
The other half area $LA_1A_2A_3$ can be spanned by inverting the sign of $\theta$ by local operations.
Thus, we can span the whole area of the Weyl chamber with 3 $R_{ZZ}(\theta)$ at most, 
which implies that the continuous basis gate $R_{ZZ}(\theta)$ can generate arbitrary SU(4) gates. 
Therefore, we focus on the pulse-efficient implementaion of $R_{ZZ}(\theta)$ on CR-based superconducting quantum devices, hereafter. 

\subsection{Hamiltonian driven 2-qubit unitary operations}
\begin{figure}[htbp]  
\centering  
\includegraphics[clip,width = 6cm]{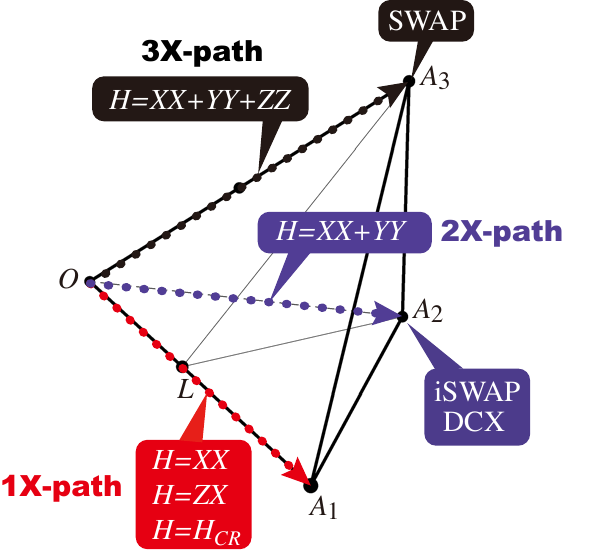}  
\caption{Native 2-qubit gates generated by various Hamiltonians. SU(4) gates on each dotted line  
can be realized as $\exp [-i H t]$ with corresponding $H$ belonging to each path (Red dotted line:1X, Purple:2X, Black:3X).  
Here, $H_{CR}$ denotes CR-Hamiltonian.}  
\label{fig:native_gates}  
\end{figure}

We consider the unitary operation driven by the time evolution of a given Hamiltonian.  
Typical cases are given in Ref.~\cite{PhysRevA.67.042313}.  
For the isotropic exchange Hamiltonian, $H = XX + YY + ZZ$, the corresponding propagator $\exp[-i H t]$ gives 
unitary operations along the axis $OA_3$ in Fig.~\ref{fig:native_gates}, which we call the 3X-path since it varies 
the 3 Cartan coefficients simultaneously.  
Likewise, $\exp[-i H t]$ with the 2D exchange Hamiltonian $H = XX + YY$ produces unitaries along the axis $OA_2$ (2X-path).  

Here, we call 2-qubit gates that can be produced by a single propagator 'native' gates,  
which are efficiently realized on the corresponding device.  
From Fig.~\ref{fig:native_gates}, we find the SWAP gate is native to the isotropic exchange Hamiltonian system,  
and so is the iSWAP gate to the 2D exchange Hamiltonian.  

Now, we consider the effective Hamiltonian, $H_{CR}$, for the cross-resonance-driven superconducting device, 
which is given as 
\cite{rigetti2010fully, chow2011simple, sheldon2016procedure, magesan2020effective}
\begin{align}
\label{eq:h_cr}
    H_{CR} &= \nu_{ZX} ZX + \nu_{ZY} ZY + \nu_{ZZ} ZZ \nonumber \\
           &+ \nu_{IX} IX + \nu_{IY} IY + \nu_{IZ} IZ + \nu_{ZI} ZI,
\end{align}
where $I$ denotes the identity operator. 
Here, we tacitly imply that the CR-pulse is applied to the first qubit.
Note that the classical crosstalk effect is represented by non-zero coefficients $\nu_{ZY}$ and $\nu_{IY}$. 
We need to diagonalize $H_{CR}$ to find the exact propagator with explicit $t$-dependence of the exponent. 
However, from the properties of the Lie bracket, it can be readily understood that the CR-Hamiltonian-driven propagator 
generates a the 2-qubit unitary operator belonging to the 1X-path of the Weyl chamber. (See Appendix~\ref{app:CR-Hamiltonian propergator})

\begin{figure}[ht]
\centering
\includegraphics[clip,width = 8.5cm]{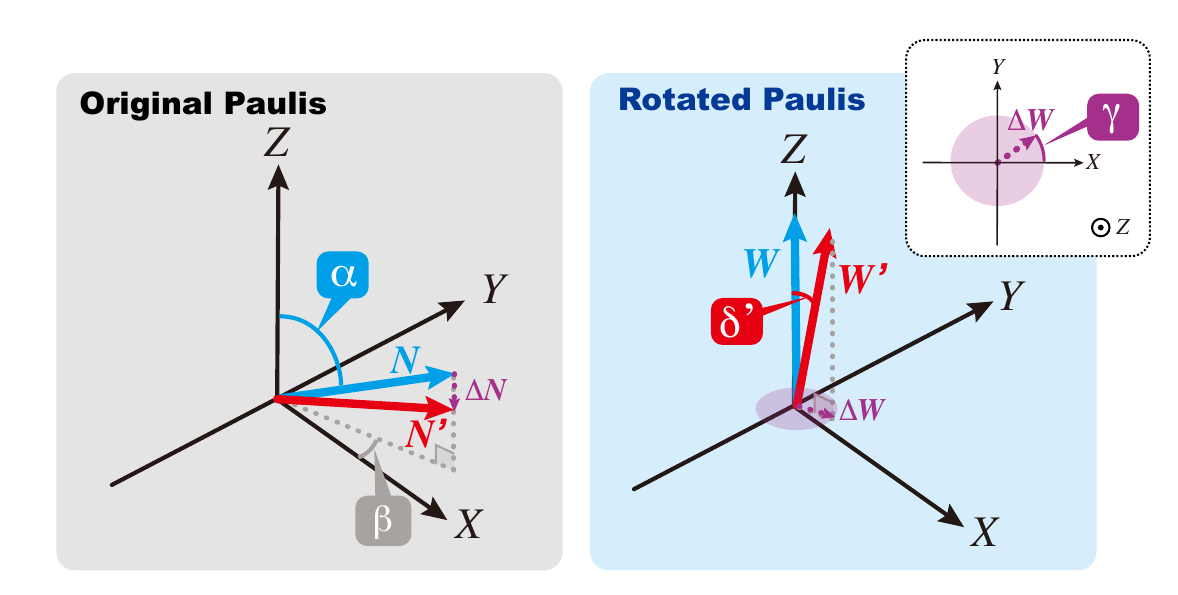}
\caption{Geometrical relation between Pauli-$N$($N'$) and rotated $W$($W'$).
Here, $\alpha$ and $\beta$ are zenith and azimuth angles of $N$, while $\delta'$ and $\gamma$ are those of $W'$.}
\label{fig:n_vector}
\end{figure}

We rewrite the CR-Hamiltonian Eq.~(\ref{eq:h_cr}) for further analysis as 
\begin{align}
\label{eq:H_cr_nn}
    H_{CR} = \nu_{ZN} ZN + \nu_{IN'} IN' + \nu_{ZI} ZI, 
\end{align}
with $\nu_{ZN} = \sqrt{|\nu_{ZX}|^2 + |\nu_{ZY}|^2 + |\nu_{ZZ}|^2}$ and $\nu_{IN'} = \sqrt{|\nu_{IX}|^2 + |\nu_{IY}|^2 + |\nu_{IZ}|^2}$. 
Here, we define general Pauli operators $N$ and $N'$ as
$N  = \bar{\nu}_{ZX} X + \bar{\nu}_{ZY} Y + \bar{\nu}_{ZZ} Z$ and 
$N' = \bar{\nu}_{IX} X + \bar{\nu}_{IY} Y + \bar{\nu}_{IZ} Z$
with $\bar{\nu}_{ZA} = \nu_{ZA} / \nu_{ZN}$ and $\bar{\nu}_{IA} = \nu_{IA} / \nu_{IN}$. 
Geometrical relation between Pauli-$N$ and $X$, $Y$, $Z$ is shown in Fig.~\ref{fig:n_vector}, 
which is characterized by zenith and azimuthal angles $\alpha$ and $\beta$. 
Note that these angles are related to the Hamiltonian coefficients as 
$(\bar{\nu}_{ZX}, \bar{\nu}_{ZY}, \bar{\nu}_{ZZ}) = (\sin\alpha \cos\beta, \sin\alpha \sin \beta, \cos\alpha)$. 

To simplify the formulation, we introduce a super-operator $\mathbb{R}(\alpha, \beta)$ to express a specific rotation on a Pauli operator $A$,  
\begin{align}
    \mathbb{R}&(\alpha, \beta)[A] \equiv \nonumber \\
    & \left\{\mathbbm{1} \otimes r_Y(\alpha) r_Z(\beta) \right\} \cdot A \cdot \left\{\mathbbm{1} \otimes r_Z(-\beta) r_Y(-\alpha) \right\}.
\end{align}
As shown in Fig.~\ref{fig:n_vector}, $\mathbb{R}(\alpha, \beta)$ consists of successive rotations, $-\beta$ around $Z$-axis, 
then $-\alpha$ around $Y$-axis. 
Here, we choose $\alpha$ and $\beta$ so as to satisfy $\mathbb{R}(\alpha, \beta)[N] \equiv W = Z$.
Then, applying $\mathbb{R}(\alpha, \beta)$ to the CR-Hamiltonian, Eq.~(\ref{eq:H_cr_nn}), gives
\begin{align}
\label{eq:}
    \tilde{H}_{CR} &\equiv \mathbb{R}(\alpha, \beta)[H_{CR}] \nonumber \\
                   &= \nu_{ZN} ZZ + \nu_{IN'} IW' + \nu_{ZI} ZI. 
\end{align}
The propagator driven by this rotated Hamiltonian, $\tilde{U}_{CR}(t) = \exp \left[-i \tilde{H}_{CR} t \right]$, 
satisfies the relation 
\begin{align}
U_{CR}(t) = \mathbb{R}^{-1}(\alpha, \beta)[\tilde{U}_{CR}(t)].
\label{eq:UCR_UtilderCR}
\end{align}
Thus, we hereafter consider $\tilde{U}_{CR}(t)$ instead of $U_{CR}(t)$.

For the special case $N \approx N'$, or $W \approx W'$, we can apply the perturbation theory. 
Considering $N' = N + \Delta N$ with small perturbation $\Delta N$, 
we split $\tilde{H}_{CR}$ as
\begin{align}
     \tilde{H}_{CR} &= H_0 + V, \\
                H_0 &= \nu_{ZN} ZZ + \nu_{IN'} IZ + \nu_{ZI} ZI, \\
                  V &= \nu_{IN'} I \Delta W,
\end{align}
where $\Delta W \equiv \mathbb{R}(\alpha, \beta)[\Delta N]$.
Since $N \approx N'$, one can consider $\Delta N$ is nearly perpendicular to $N$, 
and consequently we find $\Delta W$ in the $X$-$Y$ plane as shown in Fig.~\ref{fig:n_vector}. 
Under the idealistic condition with crosstalk-free and small $ZZ$-interaction limit, 
we expect $(\alpha, \beta) \approx (\pi/2,0)$.
Straightforward application of perturbation theory in the interaction picture \cite{messiah2014quantum}, we find
\begin{align}
\label{eq:U(t)=UI_U0}
    \tilde{U}(t) = \exp \left[-i \tilde{H}_{CR} t \right] = U_I(t) \cdot U_0(t), 
\end{align}
where
\begin{align}
\label{eq:U0=exp[-iH0t]}
    U_0(t) &= e^{-i H_0 t} \nonumber \\
           &= e^{-i \nu_{ZN} t ZZ} \cdot e^{-i \nu_{IN'} t IZ} \cdot e^{-i \nu_{ZI} t ZI}, \\
\label{eq:UI}
    U_I(t) &= 1 + \sum_{n=1}^{\infty} U_I^{(n)}(t).
\end{align}
The formal solution of $U_I^{(n)}$ is given as 
\begin{align}
\label{eq:U_I^(n)}
    &U_I^{(n)}(t) = \nonumber \\
    &(-i)^n \int_{0}^{t} dt_1 \int_{0}^{t_1} dt_2 \cdots \int_{0}^{t_{k-1}} dt_k V_I(t_1) V_I(t_2) \cdots V_I(t_k), 
\end{align}
where $V_I(t) = U_0^\dagger(t) \cdot V \cdot U_0(t)$.
We consider up to the first-order perturbation term, 
\begin{align}
\label{eq:U_I^(1)}
    U_I^{(1)}(t) &= -i \int_{0}^{t} dt_1 V_I(t_1) \nonumber \\
                 &= -i \nu_{IN'} \int_{0}^{t} dt_1 U_0^\dagger(t_1) \cdot \{I \Delta W\} \cdot U_0(t_1).
\end{align}
Here, we rewrite $\nu_{IN'} \Delta W$ as $\delta W_\gamma$, 
where $\delta = \nu_{IN'} |\Delta W| = \nu_{IN'} \sin \delta'$ 
and $W_\gamma = \cos \gamma X + \sin \gamma Y$. 
Note that $\delta'$ and $\gamma$ are the zenith and azimuth angles of $W'$ as shown in Fig.~\ref{fig:n_vector}). 
Then, we obtain
\begin{align}
\label{eq:integrand}
    U_0^\dagger&(t) \cdot \{I \Delta W\} \cdot U_0(t) \nonumber\\
    &= \cos \nu_{ZN} t \cdot \{\cos (\gamma + \nu_{IN'}t) X + \sin (\gamma + \nu_{IN'}t) Y\}.
\end{align}
Substituting Eq.~(\ref{eq:integrand}) into Eq.~(\ref{eq:U_I^(1)}) and carrying out the integration, we obtain
\begin{align}
\label{eq:deltaIW_gamma}
    U_I^{(1)}(t) =  \delta I W_\gamma(t),
\end{align}
where
\begin{align}
    W_\gamma(t) &= w_{X,\gamma}(t) X + w_{Y,\gamma}(t) Y, \label{eq:W_gamma} \\
    w_{X,\gamma}(t) &= \frac{\sin (\nu^{(+)} t + \gamma)}{2\nu^{(+)}} + \frac{\sin (\nu^{(-)} t + \gamma)}{2\nu^{(-)}} \nonumber \\
                    &- \frac{\nu_{IN'} \sin \gamma}{\nu^{(+)} \nu^{(-)}},\label{eq:w_X_gamma}\\
    w_{Y,\gamma}(t) &=-\frac{\cos (\nu^{(+)} t + \gamma)}{2\nu^{(+)}} - \frac{\cos (\nu^{(-)} t + \gamma)}{2\nu^{(-)}} \nonumber \\
                    &+ \frac{\nu_{IN'} \cos \gamma}{\nu^{(+)} \nu^{(-)}},\label{eq:w_Y_gamma}
\end{align}
with $\nu^{(\pm)}= \nu_{ZN} \pm \nu_{IN'}$.
The overall unitary $\tilde{U}_{CR}(t)$ up to the first order perturbation gives
\begin{align}
    \tilde{U}_{CR}(t) &= (1 + U_I^{(1)}(t)) \cdot U_0(t) \nonumber \\
         &= (1 -i \delta I W_\gamma(t)) \cdot U_0(t) \nonumber \\
         & \approx e^{-i \delta I W_\gamma(t)} \cdot U_0(t) \equiv U_1(t) \cdot U_0(t).
\end{align}
We define $U_1(t) \equiv e^{-i \delta I W_\gamma(t)}$ presuming $\delta |W_\gamma(t)| \ll 1$.
Then, we finally obtain 
\begin{align}
\label{eq:U_CR_tilder}
    \tilde{U}_{CR}(t) = e^{-i \delta I W_\gamma(t)} \cdot e^{-i \nu_{ZN} t ZZ} \cdot e^{-i \nu_{IN'} t IZ} \cdot e^{-i \nu_{ZI} t ZI}.
\end{align}
One can see from Eq.~(\ref{eq:U_CR_tilder}) that the only non-zero Cartan coefficient, $c$, 
is linearly dependent to $t$, i.e., $c=\nu_{ZN}t$. 
We numerically investigate the $\delta$-dependence of the CR-Hamiltonian-driven unitary and identify the parameter region 
in which the above first-order perturbation picture remains valid. 

\subsection{Unitary process tomography (UPT)}
Designing a 2-qubit gate on real quantum devices requires evaluating the fidelity of the experimentally designed gate compared to the target ideal unitary.
Quantum process tomography (QPT) is often considered the standard approach for this purpose 
~\cite{mohseni2008quantum, bialczak2010quantum}. 
However, full QPT for a 2-qubit system requires 144 experiments to characterize the quantum channel, including its non-unitary components.
We aim to reduce the number of required experiments by considering prior information. 
Assuming the unitarity of the target quantum channel, we can employ minimal probe and measurement sets~\cite{PhysRevA.90.012110, gutoski2014process}.
A set of probe states is termed unitarily informationally complete (UIC) if it provides sufficient information to distinguish any two unitary maps. 
The UIC set composed of pure states for 2-qubit system is given as 
$\{\ket{00}, \ket{01}, \ket{10}, \ket{++}\}$, where $\ket{++} = 1/2(\ket{0}+\ket{1})\otimes(\ket{0}+\ket{1})$~\cite{PhysRevA.90.012110}.
Thus, we apply the unknown near-unitary quantum channel to these four initial states and perform state tomography (ST) to characterize the output states.
Since the general ST of a 2-qubit system requires 9 experiments with Pauli Z measurements, 
we need to conduct 36 experiments to identify the unknown unitary. 
Note that the number is significantly reduced compared to the 144 required for full QPT. 

Here is the procedure for unitary process tomography (UPT) of an unknown near-unitary quantum channel $\mathbb{G}$
\begin{enumerate}
    \item Introduce a parameterized unitary operator $U(\Theta)$ with the parameter set $\Theta$.
    \item Apply $\mathbb{G}$ to the UIC set $\{\ket{u_i}\} (i=0,1,2,3) \equiv \{\ket{00}, \ket{01}, \ket{10}, \ket{++}\}$
and perform state tomography (ST) to obtain $\rho_i = \mathbb{G}[\ket{u_i}\bra{u_i}]$.
    \item Find optimal parameter set $\Theta^*$ to maximize the fitness  $F(\Theta) \equiv |\frac{1}{4} \sum_{i=0}^3 \mathrm{tr} [\rho_i \sigma_i(\Theta)]|^2$, 
where $\sigma_i(\Theta) = U(\Theta) \ket{u_i} \bra{u_i} U^\dagger(\Theta)$.
    \item Obtain the unitary operator $U(\Theta^*)$, which approximates the near-unitary quantum channel $\mathbb{G}$.
\end{enumerate}

We define three types of errors, $\varepsilon_f, \varepsilon_p, \varepsilon_u$, corresponding to fitness, purity, and unitarity errors, as 
\begin{align}
\label{eq:UPT_errors}
    \varepsilon_f &= 1 - F(\Theta^*), \nonumber \\
    \varepsilon_p &= \frac{1}{4} \sum_{i=0}^3 (1 - \mathrm{tr}\left[\rho_i \rho_i \right]), \nonumber \\
    \varepsilon_u &= \frac{1}{3} \sum_{i=0}^2 \left| \frac{1}{2} - \mathrm{tr}\left[\rho_i \rho_3\right] \right|.
\end{align}

Since we are interested in the UPT of CR-Hamiltonian-driven propagator, $\exp\left[-i H_{CR} t\right]$, 
we introduce parameterized unitary based on Eq.~(\ref{eq:U_CR_tilder}) together with Eq.~(\ref{eq:UCR_UtilderCR})
as
\begin{align}
\label{eq:U_Theta_t}
    U(\Theta, t) &\equiv U_{CR}(t) = \mathbb{R}^{-1}(\alpha, \beta)[\nonumber \\ 
                 &e^{-i \delta I W_\gamma(t)} \cdot e^{ -i (\nu_{ZN} t + \phi_{ZZ}) ZZ} \cdot \nonumber \\ 
                 &e^{-i IZ (\nu_{IN'} t + \phi_{IZ})} \cdot e^{ -i ZI (\nu_{ZI} t + \phi_{ZI})}], 
\end{align}
where $\Theta \equiv (\alpha, \beta, \gamma, \delta, \nu_{ZN}, \nu_{IN'}, \nu_{ZI}, \phi_{ZZ}, \phi_{IZ}, \phi_{ZI})$. 
In addition to the parameters originating from the CR-Hamiltonian,
we introduce effective initial phase parameters $\phi_{ZZ}, \phi_{IZ}, \phi_{ZI}$, 
which arise from the transient dynamics induced by the rising and falling edges of the CR-pulse envelope.

\section{Results}\label{sec:Results}
\subsection{Experimental settings}
\begin{figure}[ht]
\centering
\includegraphics[clip, width=8.6cm]{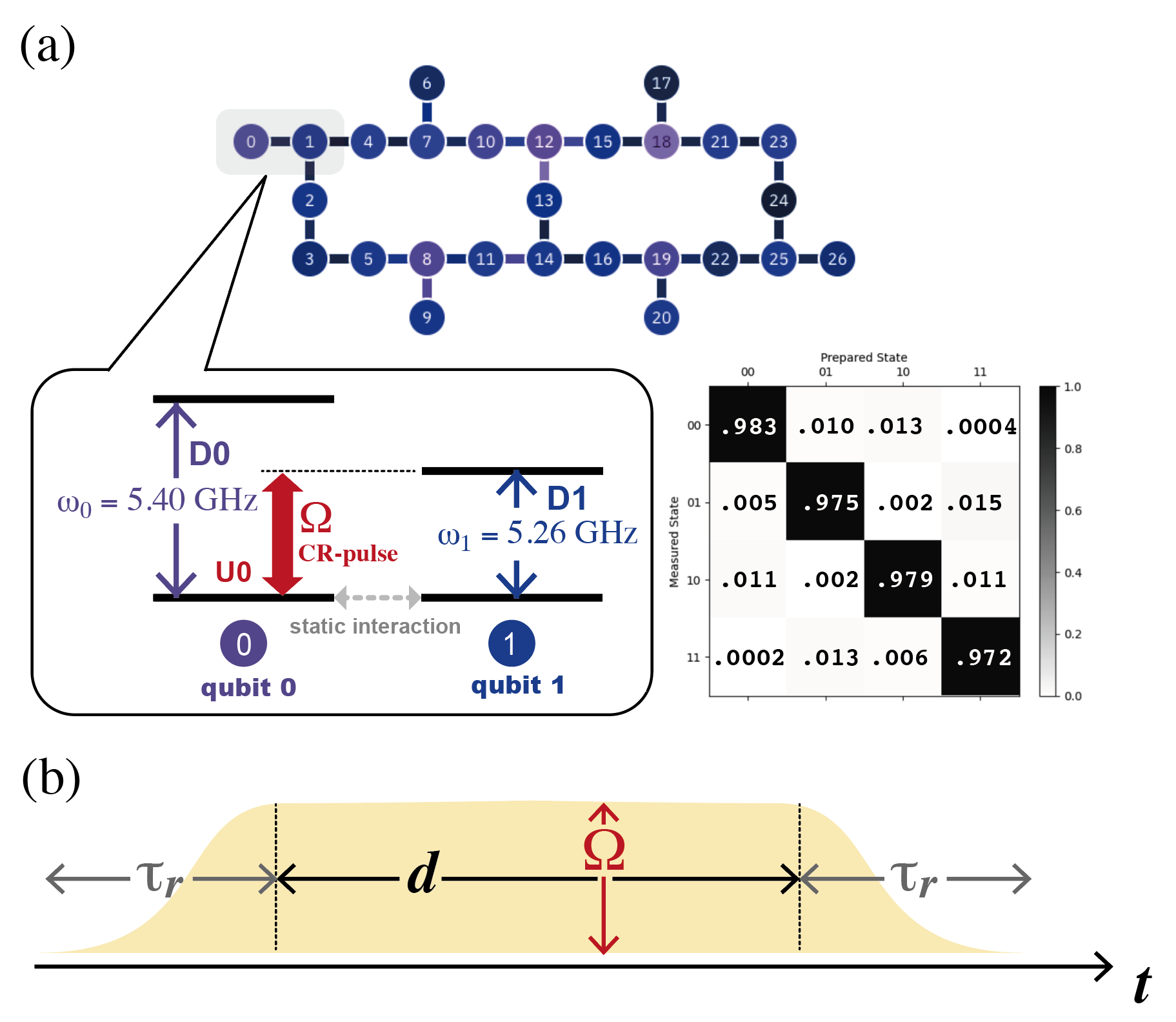}
\caption{
(a) Qubit connectivity of {\tt ibm\_kawasaki} and the readout mitigation matrix for qubits 0 and 1.
The drive channels D0 and D1 are assigned to the resonant local transitions $\omega_0=5.40$GHz and $\omega_1=5.26$GHz for qubits 0 and 1, respectively.
The control channel U0 corresponds to the CR-pulse with $\omega_1$, applied to qubit 0.
(b) Gaussian-square pulse envelope and associated parameters. $\Omega$, $d$, and $\tau_r$ denote pulse amplitude, duration, and 
steepness parameter for the rising and falling shape of the pulse envelope.}
\label{fig:ibm_kawasaki}
\end{figure}

We carried out all the experiments using qubits 0 and 1 of the IBM Quantum's 27-qubits device, {\tt ibm\_kawasaki}. 
Transition frequencies of those qubits and that of the cross-resonance pulse are depicted in Fig.~\ref{fig:ibm_kawasaki}(a). 
We obtained the read-out mitigation matrix for qubits 0 and 1, as shown in Fig.~\ref{fig:ibm_kawasaki}(a), 
and applied it to all the experimental results
~\cite{bravyi2021mitigating}.

For the CR-pulse envelope, we adopt the Gaussian-square function defined as 
\begin{align}
\label{eq:gaussian_square}
    f(t) = 
    \begin{cases}
       \Omega \exp \left[ -\frac{(t - \tau_r)^2}{\sigma^2} \right] - f_0,   & (0 \leq t < \tau_r) \\
       \Omega  - f_0,                                                       & (\tau_r \leq t < \tau_r + d) \\
       \Omega \exp \left[ -\frac{(t -\tau_r-d))^2}{\sigma^2} \right] - f_0, & (\tau_r + d \leq t < 2 \tau_r + d)
  \end{cases}
\end{align}
where $f_0 = \exp \left[ -\tau_r^2/\sigma^2 \right]$ and $\Omega$ denotes the pulse amplitude.
As shown in Fig.~\ref{fig:ibm_kawasaki}(b), the total pulse duration is $2 \tau_r + d$, where $d$ represents the duration of the constant amplitude $\Omega$.

A 2-qubit gate implemented with a single Gaussian-square pulse on the U0 channel is referred to as a crude CR-pulse gate, 
corresponding to the propagator $U_{CR}(d) \equiv \exp[-i H_{CR} d]$.
Note that $d$ is treated as $t$ in Sec.~\ref{sec:Preliminaries}, as the effective Hamiltonian $H_{CR}$ is valid only during the constant amplitude $\Omega$ portion of the pulse.
The pulse duration $d$ is quantized in discrete steps defined by the device-specific time step $dt$, set to $dt=0.222$~ns for \texttt{ibm\_kawasaki}. 
Consequently, $dt$ is used as the time unit, and $d$ is represented as a dimensionless quantity.
Other detailed device information is shown in Appendix~\ref{app:device_info} together with the version numbers of the Qiskit library used in the present work.

\subsection{Unitary process tomography of crude CR-pulse gate}
\begin{figure}[ht]
\centering
\includegraphics[clip, width=8.6cm]{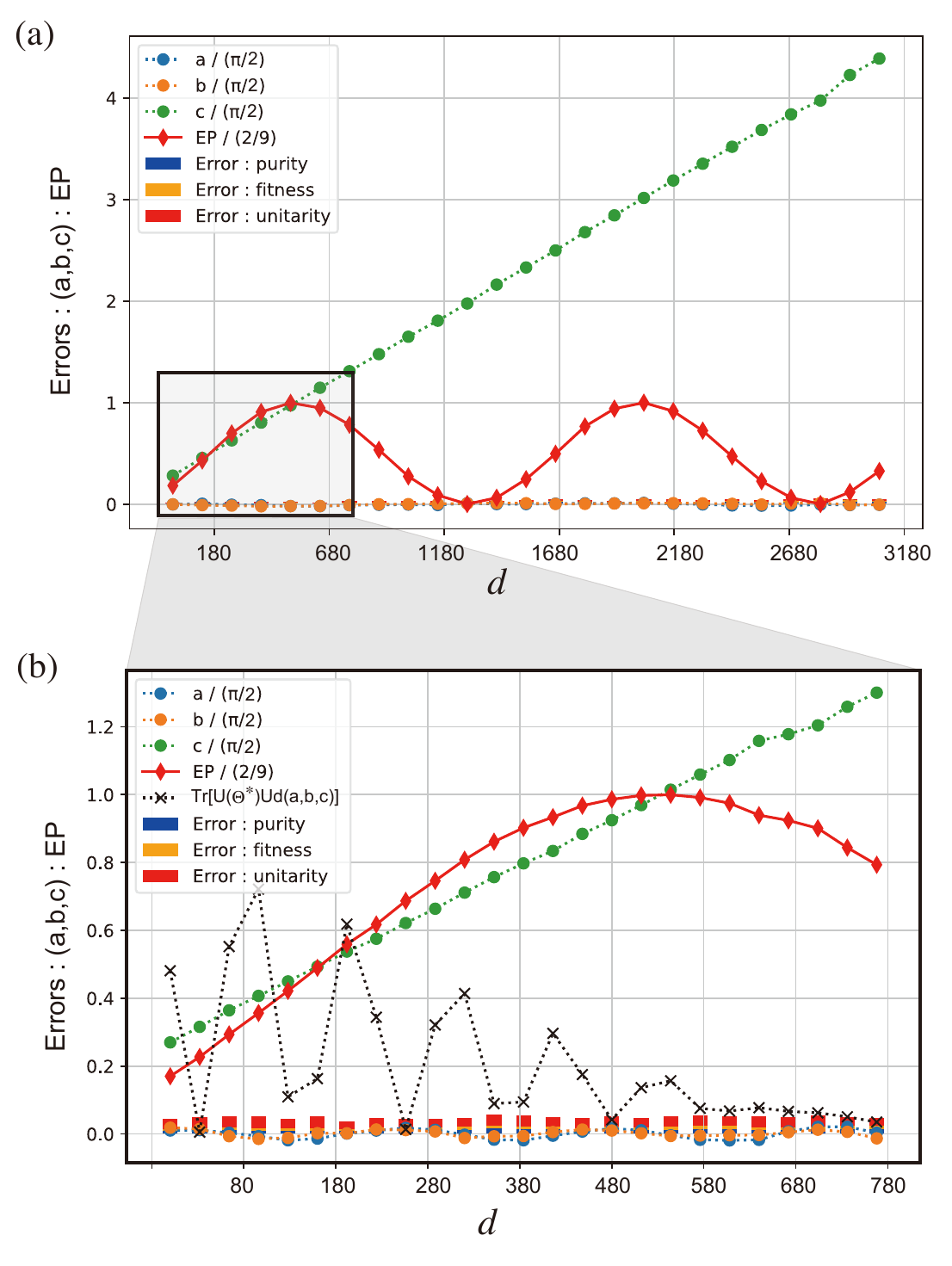}
\caption{UTM results for the crude CR-pulse gate varying pulse duration $d$.
(a) UTM results over the range $0 \leq d \leq 3073$ with a step size of 128.
(b) UTM results over the range $0 \leq d \leq 769$ with a step size of 32, corresponding to the inset region in (a).
Blue, orange, and green dots with dashed lines represent the Cartan coefficients $a$, $b$, and $c$, each normalized by $\pi/2$.
The red line denotes the EP, normalized by $2/9$.
The black dashed line represents the trace fidelity between the identified unitary $U(\Theta^*)$ and $U_d(0,0,c)$.
The blue, orange, and red bars in the chart correspond to the errors $\varepsilon_p$, $\varepsilon_f$, $\varepsilon_u$, respectively.
}
\label{fig:utm_result_amp06}
\end{figure}
\begin{table*}[ht]
\caption{Parameters obtained from the UPT fitting to $U(\Theta)$ for the crude CR-pulse gate on qubits (0,1) of {\tt ibm\_kawasaki}.
The values for $\nu_{ZN}$, $\nu_{IN'}$, and $\nu_{ZI}$ are expressed in the unit of frequency, $dt^{-1}=4.5\text{ns}^{-1}$.
}
\label{tab:U_H_parameters}
\begin{tabular}{c c c c c c c c c c}
\hline
\hline
$\alpha$    & $\beta$       & $\gamma$      & $\delta$      & $\nu_{ZN}$            & $\nu_{IN'}$           & $\nu_{ZI}$           & $\phi_{ZZ}$            & $\phi_{IZ}$           & $\phi_{ZI}$\\
\hline
$0.499 \pi$ & $0.042 \pi$   & $0.072 \pi$   & $0.024 \pi$  & $1.06 \times 10^{-3}$ & $1.57 \times 10^{-3}$ & $5.58 \times 10^{-2}$ & $4.04\pi \times 10^{-3}$  & $1.38\pi \times 10^{-2}$ & $0.62\pi$\\
\hline
\hline
\end{tabular}
\end{table*}
We applied the UPT to a crude CR-pulse gate and identified the corresponding unitary operator as a function of the duration $d$. 
The parameters for the Gaussian-square envelope are set to $\Omega=0.6$, $\tau_r=160$, and $\sigma=80$. 
Since 900 experiments can be loaded into a single job on {\tt ibm\_kawasaki}, we conducted 25 UTM experiments, each with 8192 shots, to generate Fig.~\ref{fig:utm_result_amp06}.

Initially, we varied the duration parameter $d$ from 0 to 3073 in steps of 128 to observe the long-time behavior of the crude CR-pulse gate as shown in Fig.~\ref{fig:utm_result_amp06}(a). 
Furthermore, a similar experiment was conducted to closely examine the range of $d$ corresponding to the EP range $0<\mathrm{EP}<2/9$, which is required for generating the SU(4) gate (See Fig.~\ref{fig:utm_result_amp06}(b)).
All the parameters $\Theta$ determined by the UPT is listed in the Table~\ref{tab:U_H_parameters} and the associated errors, $\varepsilon_f, \varepsilon_p, \varepsilon_u$ are shown in Fig.~Fig.~\ref{fig:utm_result_amp06}.
As the Cartan decomposition is not uniquely determined, we selected the decomposition such that the Cartan coefficients satisfy $|c|\geq|b|\geq|a|$. 
Note that the only Cartan coefficient $c$ takes non-zero values over the scan range of duration $d$, while $b \approx a \approx 0$.
The EP value reaches its maximum of 2/9 at $d=544$, indicating that this crude CR-pulse gate has the potential to produce CX, CZ, $R_{ZZ}(\pi/2)$, and other 2-qubit gates in the class $[\pi/2, 0, 0]$. 

Another significant finding from Fig.~\ref{fig:utm_result_amp06} is that $c$ value exhibits distinct linear dependence to the pulse duration $d$, 
which suggests that the present crude CR-pulse gate can be expressed as the unitary operator $U(\Theta, d)$ in Eq.~(\ref{eq:U_Theta_t}). 
This demonstrates that the assumptions made in deriving Eq.~(\ref{eq:U_Theta_t}) are appropriate.

\subsection{Extraction of $\boldsymbol{R_{ZZ}(\theta)}$ from the crude CR-pulse gate}
The trace fidelity of the identified unitary $U(\Theta^*)$ with respect to $U(0,0,c)$ exhibits oscillatory behavior, which is shown as the broken line in Fig.~\ref{fig:utm_result_amp06}(b).
Thus, we need to extract $R_{ZZ}(\theta)$ to use as a basis gate. 
Since we obtain $U(\Theta^*, d)$ with optimized parameters $\Theta^*$ for crude CR-pulse gate, basis gate $R_{ZZ}(\theta)$ can be obtained as follows.  
From Eq.~(\ref{eq:U_Theta_t}), we readily find 
\begin{align}
\label{eq:exp_ZZ}
    R_{ZZ}(\theta) &= e^{-i ZZ (\nu_{ZN} d + \phi_{ZZ})} \nonumber \\
                   &= e^{i ZI (\nu_{ZI} d + \phi_{ZI})} \cdot e^{i IZ (\nu_{IN'} d + \phi_{IZ})} \cdot e^{i \delta I W_\gamma(d)} \cdot \nonumber \\
                   &\,\,\,\,\,\,\,\mathbb{R}(\alpha, \beta)[U(\Theta, d)].
\end{align}
Thus, $R_{ZZ}(\theta)$ can be extracted by setting the CR-pulse duration to $d = (\theta/2 - \phi_{ZZ})/\nu_{ZN}$, 
which satisfies the condition $\nu_{ZN} d + \phi_{ZZ} = \theta/2$.

\begin{figure*}[ht]
\begin{minipage}{\hsize}
\hspace{16cm}

\begin{quantikz}[thick]
 \qw & \gate[wires=2]{R_{ZZ}(\theta)} & \qw\\
 \qw &                                & \qw 
\end{quantikz}
=
\begin{quantikz}[thick]
  &\qw                 & \qw                & \gate[wires=2]{U_{CR}(d)} & \qw               & \qw                & \qw                & \gate{r_Z(\theta_{ZI})} & \qw \\
  &\gate{r_Y(-\alpha)} & \gate{r_Z(-\beta)} &                           & \gate{r_Z(\beta)} & \gate{r_Y(\alpha)} & \gate{r_\gamma(\delta)} & \gate{r_Z(\theta_{IZ})} & \qw 
\end{quantikz}

\vspace{0.5cm}
\end{minipage}
\caption{Quantum circuit for extracting $R_{ZZ}(\theta)$ basis gate. 
$U_{CR}(d)$ denotes a crude CR-pulse gate with the pulse duration $d=(\theta/2 - \phi_{ZZ})/\nu_{ZN}$. 
The rotation angles, $\theta_{ZI}$ and $\theta_{IZ}$, are given as $\theta_{ZI}=\nu_{ZI} d + \phi_{ZI}$ and $\theta_{IZ}=\nu_{IZ} d + \phi_{IZ}$. 
}
\label{fig:circuit_exp_ZZ}
\end{figure*}
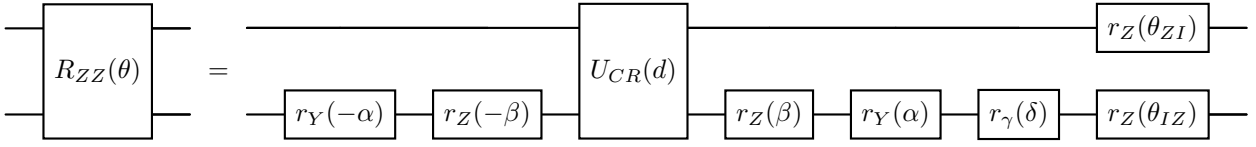

\begin{figure*}[ht]
(a) \hfill \hfill (b) \hfill \hfill \hfill 

\includegraphics[clip, width=8.8cm]{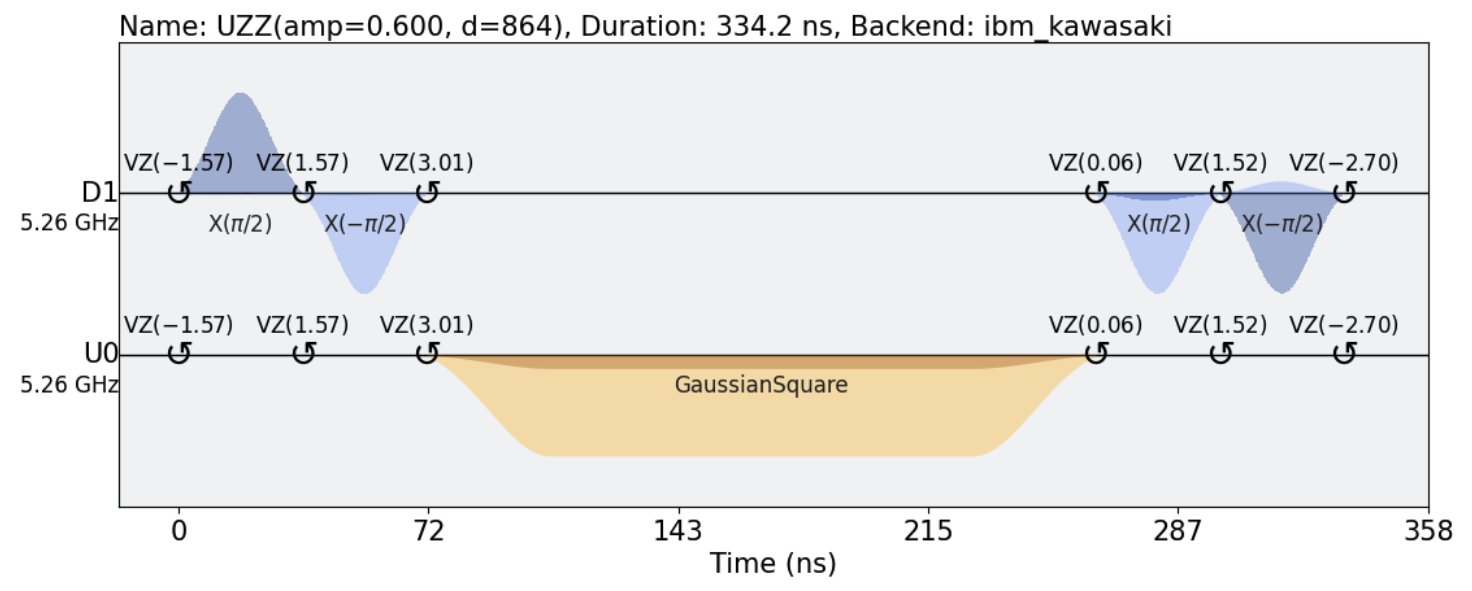}
\includegraphics[clip, width=8.8cm]{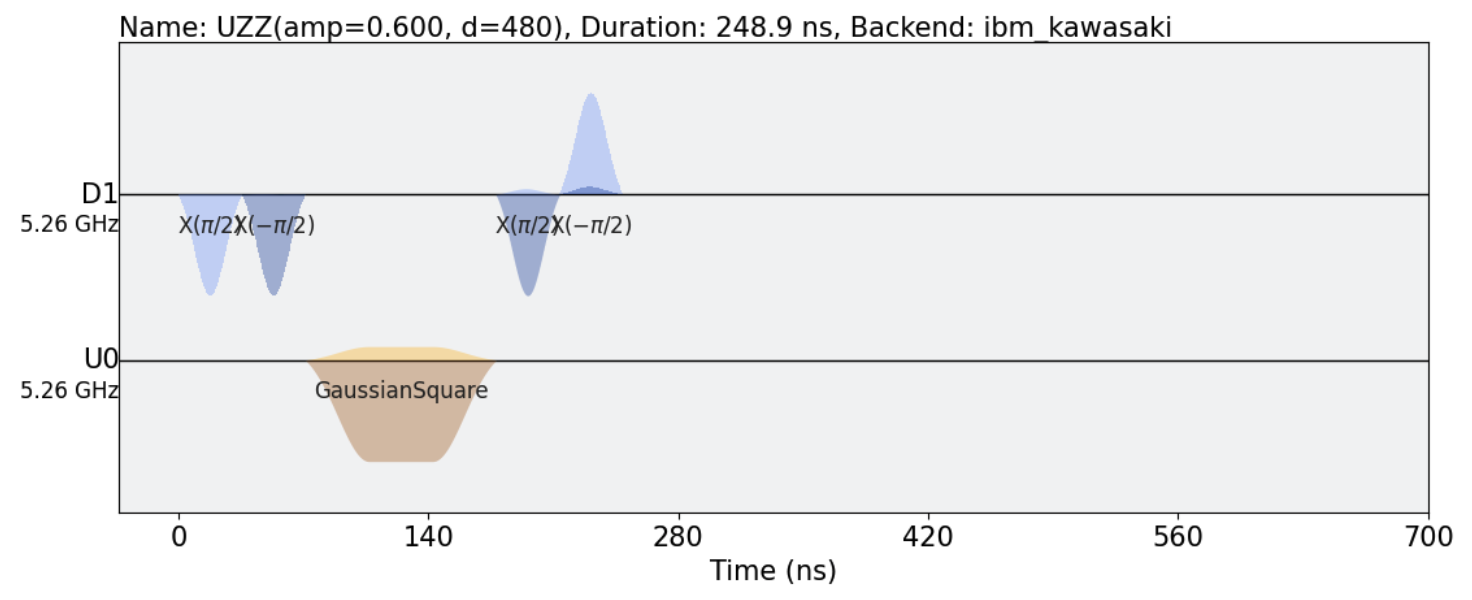}

(c) \hfill \hfill (d) \hfill \hfill \hfill 

\includegraphics[clip, width=8.8cm]{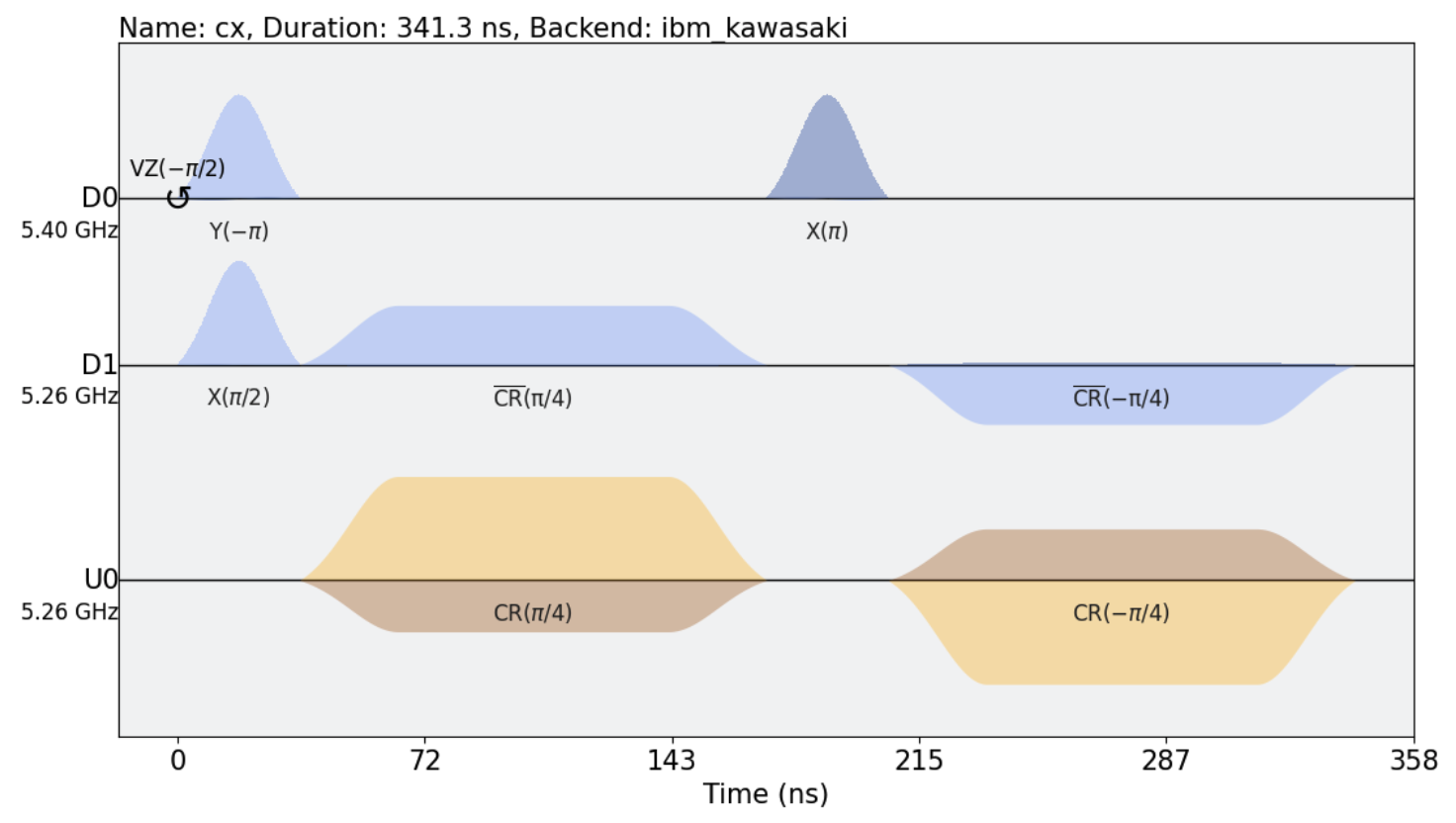}
\includegraphics[clip, width=8.8cm]{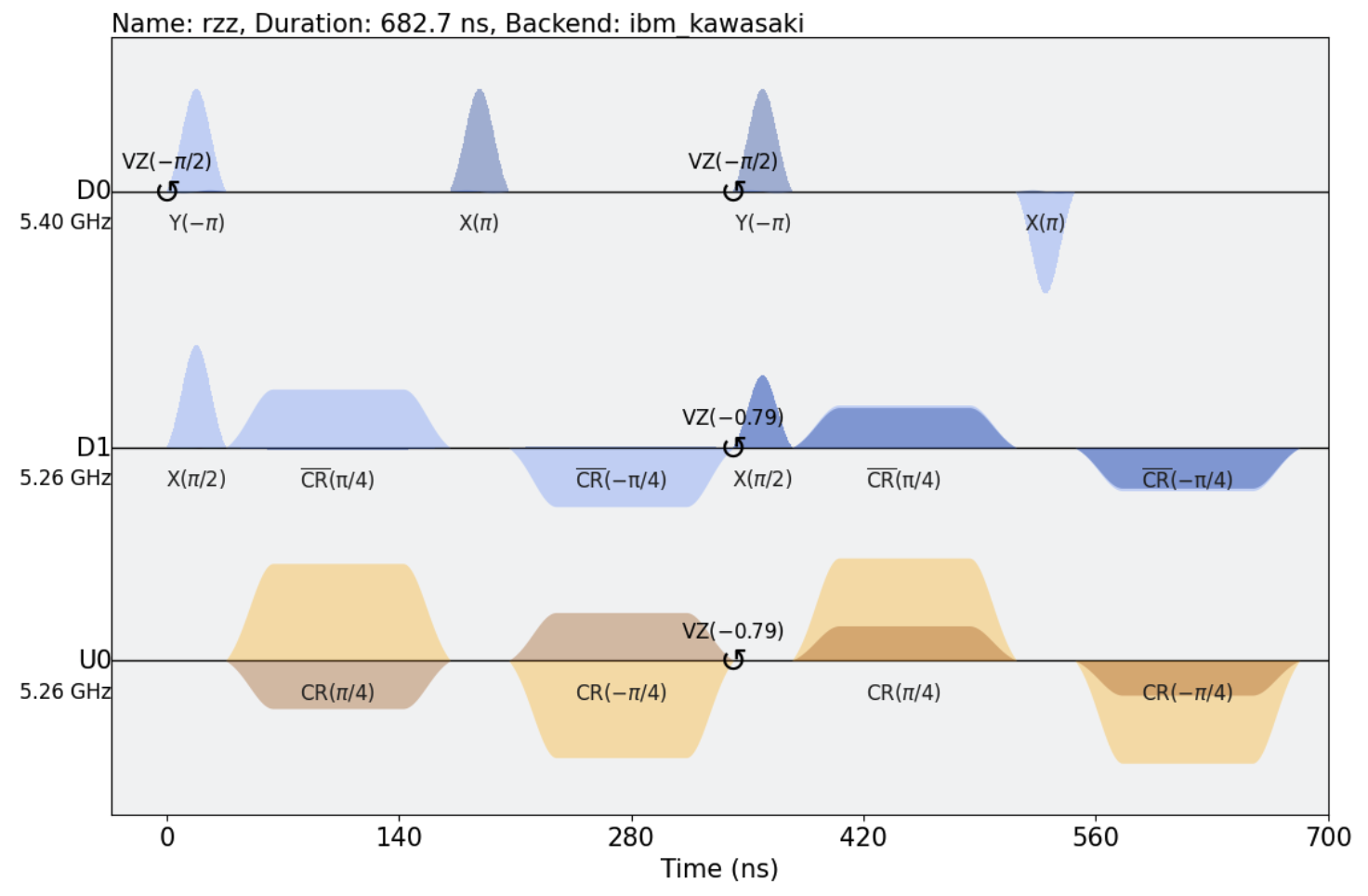}
\caption{Comparison of pulse schedules for various basis gates: 
(a) Pulse schedule of $R_{ZZ}(\pi/2)$ on qubits (0,1) realized by the circuit shown in Fig.~\ref{fig:circuit_exp_ZZ}. 
The CR-pulse envelope has an amplitude of $\Omega = 0.6$ and a duration of $d = 864 dt = 192$ns.
(b) Pulse schedule of $R_{ZZ}(\pi/4)$ on qubits (0,1). 
The CR-pulse envelope has an amplitude of $\Omega = 0.6$ and a duration of $d = 480 dt = 107$ ns.
(c) Pulse schedule of {\tt ibm\_kawasaki}'s default CX basis gate on qubits (0,1).
(d) Pulse schedule of $R_{ZZ}(\pi/4)$ with default 2 CX decomposition on qubits (0,1).
}
\label{fig:pulse_schedules}
\end{figure*}

Figure~\ref{fig:circuit_exp_ZZ} illustrates the quantum circuit corresponding to the extraction process described in Eq.~(\ref{eq:exp_ZZ}). 
Here, we define the local gate $r_\gamma(\delta) \equiv e^{-i \delta W_\gamma(d)}$. 
Successive $r_Z$, $r_Y$, and $r_\gamma$ gates before and after the crude CR-pulse gate can be combined into a single local operation. 
As a result, the overall pulse schedule for $R_{ZZ}(\pi/2)$ is simplified, as shown in Fig.~\ref{fig:pulse_schedules}(a).
No pulses are applied to the drive channel D0, as the counter phase gate $r_Z(\theta_{ZI})$ is implemented via the virtual-$Z$ gate on IBM Quantum devices~\cite{mckay2017efficient}.
This pulse-less feature on the D0 channel contrasts with the echo-based CX gate, as shown in Fig.~\ref{fig:pulse_schedules}(c), 
where the control bit is flipped and re-flipped using X-gate operations on the D0 channel.
In addition to that, active canceling pulses are also applied on D1 channel in Fig.~\ref{fig:pulse_schedules}(c). 
In contrast, no pulse is applied to the D1 channel during CR-pulse irradiation in Fig.~\ref{fig:pulse_schedules}(a).

Shown in Fig.~\ref{fig:pulse_schedules}(b) is the pulse schedule for $R_{ZZ}(\pi/4)$, which can be a seed gate for implementing 
ctrl-$\sqrt{\mathrm{X}}$, $\sqrt{\mathrm{iSWAP}}$ and $\sqrt{\mathrm{SWAP}}$ gates (See Fig.~\ref{fig:Weyl-Chamber}). 
For comparison, Fig.~\ref{fig:pulse_schedules}(d) shows the pulse schedule of the default $R_{ZZ}(\pi/4)$ implementation on {\tt ibm\_kawasaki} using a 2-CX decomposition.
Because the pulse duration $d$ of the crude-CR gate is adjusted to produce the required 2-qubit interaction in Fig.~\ref{fig:pulse_schedules}(b), 
the total gate time is significantly saved compared to that of Fig.~\ref{fig:pulse_schedules}(d).

\subsection{Comparison between $\boldsymbol{R_{ZZ}(\theta)}$-based and echoed CX-based SU(4) gates}
\begin{figure*}[ht]
\centering
\includegraphics[width=17cm]{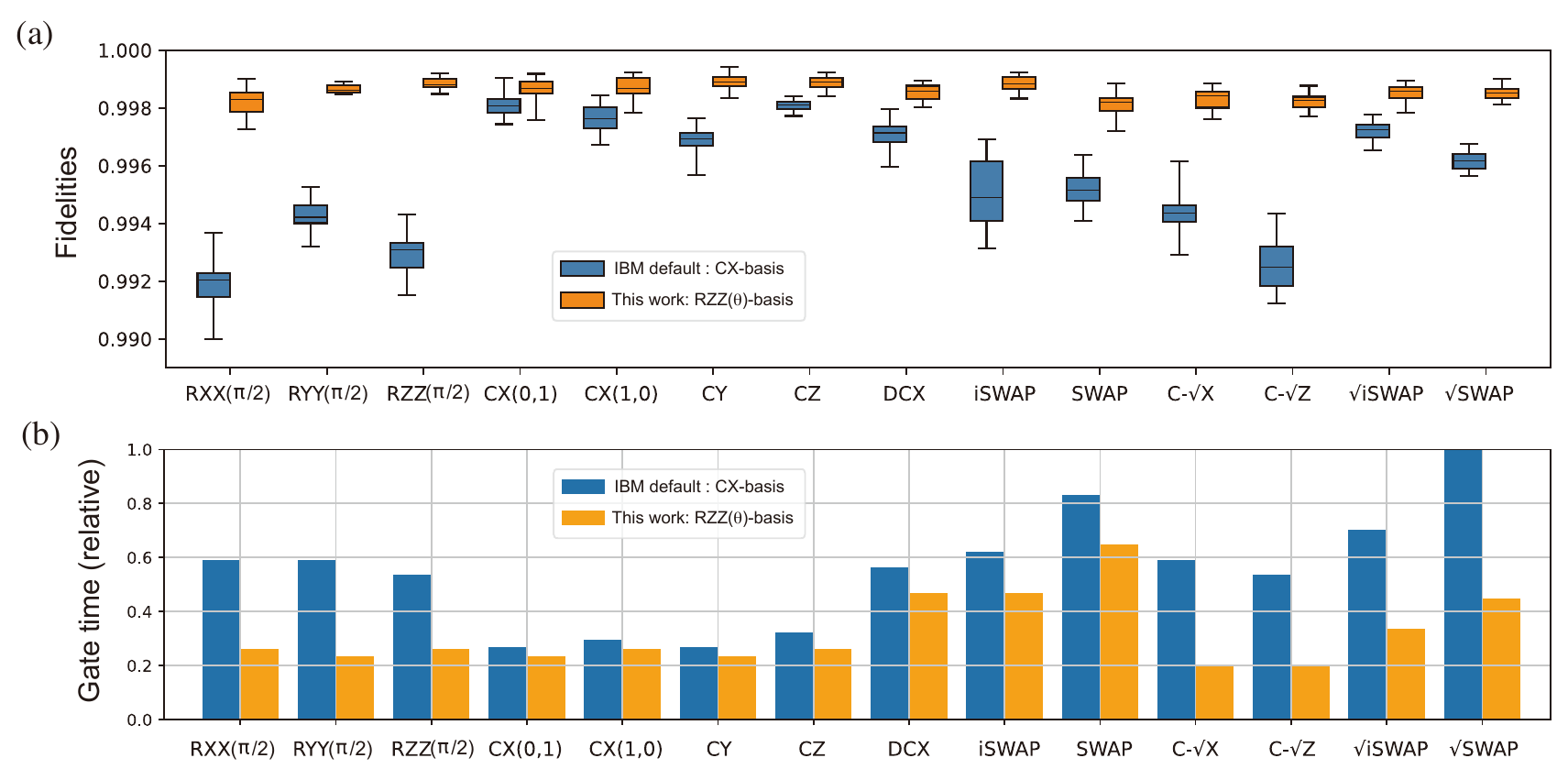}
\caption{Fidelity and gate time comparison of CX-based ({\tt ibm\_kawasaki}'s default) and $R_{ZZ}(\theta)$-based (this work) implementations for various SU(4) gates.
${\rm CX}(i,j)$ denotes CX gate in which control and target qubits are assigned to qubits $i$ and $j$, respectively. 
(a) Boxplot of 25 fidelity values measured independently. 
(b) Gate time for various SU(4) gates. The relative value is defined by taking the gate time of $\sqrt{\mathrm{SWAP}}$ gate decomposed with CX-basis ($5728 dt=1273$ns) as 1.
}
\label{fig:fidelity_gatetime}
\end{figure*}

We implemented various SU(4) gates using the proposed $R_{ZZ}(\theta)$-basis and compared their performance with the default CX-basis implementation.
Figure~\ref{fig:fidelity_gatetime} shows the gate time and fidelity comparisons for various SU(4) gates implemented with different basis gates. 
Note that the fidelity values could exceed the coherence limit because the observed histogram could be over-mitigated by the R/O-error mitigation process. 
Figure~\ref{fig:fidelity_gatetime}(a) shows that the $R_{ZZ}(\theta)$-based implementation achieves higher fidelity than the default echoed CX-based implementation across all SU(4) gates.
Since the CX gate is the sole 2-qubit basis gate on {\tt ibm\_kawasaki}, it is extensively tuned under the echo-based scheme, resulting in the highest fidelity for CX(0,1) among all derivative gates.
Figure~\ref{fig:fidelity_gatetime}(b) exhibits that the present gate implementation 
succeeds in achieving shorter gate time for all gates and significantly for $R_{XX}(\pi/2)$, $R_{YY}(\pi/2)$, $R_{ZZ}(\pi/2)$, 
ctrl-$\sqrt{\mathrm{X}}$, ctrl-$\sqrt{\mathrm{Z}}$, $\sqrt{\mathrm{iSWAP}}$, $\sqrt{\mathrm{SWAP}}$.
This improvement is primarily due to the flexibility of the continuous basis gate $R_{ZZ}(\theta)$, 
which allows precise adjustment of the CR-pulse duration to minimize the total gate time.
In contrast, IBM Quantum's default circuit transpilation decomposes these gates using 2 or 3 CX gates, leading to longer gate times.

\section{Discussions}\label{sec:Discussions}
As mentioned in Sec.~\ref{sec:Preliminaries}, the present method requires determining all the Hamiltonian origin parameters $\Theta^*$.
However, the absolute values of these parameters are not uniform as shown in Table~\ref{tab:U_H_parameters}.
Thus, it is difficult to finely determine them through a single fitting procedure to the UPT result shown in Fig.~\ref{fig:utm_result_amp06}.
This is because the primary $d$-dependence feature of the result is highly sensitive to specific parameters, 
including $\alpha$, $\beta$, $\nu_{ZN}$, $\nu_{IN'}$, $\nu_{ZI}$, $\phi_{ZZ}$, $\phi_{IZ}$, and $\phi_{ZI}$.
To finely determine other parameters, 
$\gamma$ and $\delta$, 
we need to conduct additional experiments in which the impact of these parameters is selectively amplified in the results.
See Appendix~\ref{app:fine_tomography} for the details of such experiments. 

As shown in Fig.~\ref{fig:utm_result_amp06}(b), the weak interaction gate $U_d(0,0,c)$ for $c < 0.1 \pi$ 
cannot be generated from a single crude Gaussian-square CR-pulse gate with $\Omega = 0.6$, 
since the shortest overall pulse duration is 320, that is $d+2\tau_r$ with $(d,\tau_r)=(0, 160)$. 
To generate such a weak interaction gate, two potential approaches can be considered. 
The first option is employing a CR-pulse gate with a smaller amplitude $\Omega$ than 0.6. 
However, this approach requires the whole UPT analysis to identify the parameter set $\Theta^*$ for new $\Omega$ that is resource-intensive.
An alternate practical way is to incorporate the decomposition scheme using two $[0,0,c]$ gates with $c \approx \pi/8$  
akin to the CX-decomposition.
This feature is formerly pointed out in the previous work \cite{satoh2022pulse}.
Note also that the gate time advantage is prominent for the SU(4) gates in the vicinity of the origin of the Weyl chamber.

Another advantage of the present method is the gate time reducing effect due to the simple structure of 
the implemented pulse schedule, i.e., an arbitrary SU(4) gate can be constructed with $[n {\rm CR}, (2n+2) {\rm LP}]$-pulses with $n=1 \sim 3$ 
as shown in Fig.~\ref{fig:concept}. 
Thus, any of the local gates adjacent to SU(4) gates can be merged to the local gates, $\{r_i\}$, 
which results in saving the total gate time of the deep circuit. 

As shown in Table~\ref{tab:U_H_parameters}, $\nu_{ZI}$ value is 50 times larger compared to $\nu_{ZN}$.
We find this rapid local $Z$-rotation is quite sensitive to the environment. 
In the present method, we cancel this $Z$-rotation by applying the counter-rotation, $r_Z(\theta)$, 
onto the control qubit using the information from the UPT.
Consequently, our approach may necessitate frequent re-calibration using UPT to accommodate the fluctuation of $\nu_{ZI}$ values.
In contrast, the echoed CR-pulses physically cancel out this $Z$-rotation, which might endure the fluctuation of $\nu_{ZI}$.  

We should be careful with the choice of amplitude parameter $\Omega$ value because too intense CR-pulse might induce 
undesired non-unitary process. 
The cause of this issue remains unclear, but it is likely driven by population leakage to the excited states.
We have monitored this process by checking $\varepsilon_u$ values and found that crude CR-pulse gates with $\Omega > 0.7$ 
exhibits very unstable behavior.
Thus, we set the amplitude parameter as $\Omega=0.6$ in the present study.

The present method relies on the UPT of the 2-qubit system, which is considered to be isolated.
Thus, this approach is vulnerable to the systematic error that depends on the state of the adjacent qubit, 
which is physically connected with the static $ZZ$-interaction.
To suppress such spectator errors, we could utilize dynamic decoupling\cite{niu2022effects}. 

\section{Conclusion}\label{sec:Conclusion}
In this paper, we proposed a tomography-based method for implementing continuous basis gate $R_{ZZ}(\theta)$ 
on the CR-based superconducting devices.
This method offers a simple echo-free pulse schedule for arbitrary 2-qubit interaction gates.
To achieve this, we employed the UPT to identify near-unitary processes with a reduced number of experiments compared to 
the full QPT. 
We discovered that the only non-zero Cartan coefficient $c$ of a crude Gaussian-square CR-pulse gate, which is linear to the pulse duration.
This characteristic is elucidated with the first-order time-dependent perturbation theory.
From the information obtained by the UPT, we extracted the basis gate $R_{ZZ}(\theta)$,
implemented various SU(4) gates, and compared them to the default echoed CX decomposition in fidelity and gate time.
The results demonstrated that the current approach outperforms the default implementation in terms of 
achieving higher fidelity with shorter gate times.
It is worth emphasizing that our methodology fully incorporates the static $ZZ$-interaction, 
a crucial factor that has been disregarded in the design of the echo-based implementations of 2-qubit basis gates. 

\section*{Acknowledgment}
This work was supported by the MEXT Quantum Leap Flagship Program Grant Number JPMXS0118067285 and JPMXS0120319794. 
TS is also supported by MEXT KAKENHI Grant Number 22K1978.
The author, M. S. would like to thank S. Uno for insightful advice.

\bibliography{ref.bib}
\bibliographystyle{unsrt}

\appendix
\subsection{Unitary operation as a CR-Hamiltonian propergator}\label{app:CR-Hamiltonian propergator}
Here, we examine the CR-Hamiltonian propagator,  
$U_{CR}(t) = \exp \left[-i H_{CR} t \right]$,
to find native gates for $H_{CR}$.
\begin{table}[h]
\caption{Lie brackets table for CR-Hamiltonian Pauli components.}
\label{tab:Lie-bracket}
\begin{tabular}{c|ccc|ccc|c}
                     & \textit{\textbf{ZX}} & \textit{\textbf{ZY}} & \textit{\textbf{ZZ}} & \textit{\textbf{IX}} & \textit{\textbf{IY}} & \textit{\textbf{IZ}} & \textit{\textbf{ZI}} \\ \hline
\textit{\textbf{ZX}} & 0                    & \textit{-IZ}         & \textit{IY}          & 0                    & \textit{-ZZ}         & \textit{ZY}          & 0                    \\
\textit{\textbf{ZY}} & \textit{IZ}          & 0                    & \textit{-IX}         & \textit{ZZ}          & 0                    & \textit{-ZX}         & 0                    \\
\textit{\textbf{ZZ}} & \textit{-IY}         & \textit{IX}          & 0                    & \textit{-ZY}         & \textit{ZX}          & 0                    & 0                    \\ \hline
\textit{\textbf{IX}} & 0                    & \textit{-ZZ}         & \textit{ZY}          & 0                    & \textit{-IZ}         & \textit{IY}          & 0                    \\
\textit{\textbf{IY}} & \textit{ZZ}          & 0                    & \textit{-ZX}         & \textit{IZ}          & 0                    & \textit{-IX}         & 0                    \\
\textit{\textbf{IZ}} & \textit{-ZY}         & \textit{ZX}          & 0                    & \textit{-IY}         & \textit{IX}          & 0                    & 0                    \\ \hline
\textit{\textbf{ZI}} & 0                    & 0                    & 0                    & 0                    & 0                    & 0                    & 0                   
\end{tabular}
\end{table}
Shown in Table~\ref{tab:Lie-bracket} are the Lie brackets for the Pauli operators contained in the $H_{CR}$,  
indicating that the operations of these Pauli operators are closed in the sub-group of SU(4) spanned by $\left[IX, IY, IZ, ZX, ZY, ZZ, ZI \right]$.  
Using the Zassenhaus formula
\begin{align}
\label{eq:zassenhaus}
    e^{t(A+B)} = &e^{tA} e^{tB} e^{-\frac{t^2}{2} [A,B]} \cdot \nonumber \\
                 &e^{-\frac{t^3}{6} (2[B,[A,B]] + [A,[A,B]])} \cdot e^{-\frac{t^4}{24} (\cdots)},
\end{align}
and the Baker-Campbell-Hausdorff formula
\begin{align}
\label{eq:baker_campbell}
    e^{tA} e^{tB} = e^{t (A+B) + \frac{t^2}{2}[A,B] + \frac{t^3}{12} [A,[A,B]] - \frac{t^3}{12} [B,[A,B]] + \cdots},
\end{align}
it can be shown that all the exponents of the Cartan decomposition of $\exp \left[-i H_{CR} t \right]$ contain only those 7 Pauli operators.
We need to diagonalize $H_{CR}$ to find the exact propagator with explicit $t$-dependence of the exponent. 
However, without knowing such an exact propagator, it can be concluded that the $XA, YA$ family $(A \in {X,Y,Z})$ does not mix in the exponent.
This feature leads to an important property that $U(t) = \exp \left[-i H_{CR} t \right]$ can only create the 2-qubit 
interaction along the 1X-path of the Weyl chamber. 

\subsection{Fine parameter fitting in unitary process tomography}\label{app:fine_tomography}
The main feature of results shown in Fig.~\ref{fig:utm_result_amp06} are characterized by the parameters, 
$\alpha$, $\beta$, $\nu_{ZN}$, $\nu_{IN'}$, $\nu_{ZI}$, $\phi_{ZN}$, $\phi_{IN'}$, and $\phi_{ZI}$, 
which can be efficiently determined through fitting. 
On the contrary, it is difficult to determine low-influential parameter values, $\delta$ and $\gamma$.
Thus, we conducted extra experiments in which influences of these parameters appear prominently. 

Here, we define
\begin{align}
\label{eq:Q_theta}
    Q(\theta)&= e^{i ZI (\tilde{\nu}_{ZI} \tilde{d} + \tilde{\phi}_{ZI})}
                         \cdot e^{i IZ (\tilde{\nu}_{IN'} \tilde{d} + \tilde{\phi}_{IZ})} 
                         \cdot e^{i \tilde{\delta} I W_{\tilde{\gamma}} (\tilde{d})} \nonumber \\
                        &\cdot \mathbb{R}(\tilde{\alpha}, \tilde{\beta})[U_{CR}(\tilde{d})], 
\end{align}
where $\tilde{\Theta} = \{\tilde{\alpha}, \tilde{\beta}, \tilde{\gamma}, \tilde{\delta}, \tilde{\nu}_{ZN}, \tilde{\nu}_{IN'}, \tilde{\nu}_{ZI}, 
\tilde{\phi}_{ZZ}, \tilde{\phi}_{IZ}, \tilde{\phi}_{ZI} \}$ denotes the parameter values close to the optimal set $\Theta^*$ and
$\tilde{d} = (\theta/2 - \tilde{\phi}_{ZZ})/\tilde{\nu}_{ZN}$. 
In practice, we obtain $\tilde{\Theta}$ from the fitting procedure of the results shown in Fig.~\ref{fig:utm_result_amp06}. 
We consider pseudo-identity gate $\tilde{id}(\theta, \xi)$, defined with $Q(\theta)$ as
\begin{align}
    \tilde{id}(\theta, \xi) = \bar{Q}(\theta, \xi) \cdot Q(\theta), 
\end{align}
where 
\begin{align}
    &\bar{Q}(\theta, \xi) \equiv \nonumber \\
    &\left\{\mathbbm{1} \otimes r_Y(\pi) r_Z(\xi)\right\} \cdot Q(\theta) \cdot \left\{\mathbbm{1} \otimes r_Z(-\xi) r_Y(-\pi) \right\}.
\end{align}
Note that $Q(\theta) = R_{ZZ}(\theta)$ and $\bar{Q}(\theta, \xi) = R_{ZZ}(-\theta)$ when $\tilde{\Theta} = \Theta^*$, and thus,  
$\tilde{id}(\theta, \xi)$ becomes the exact identity gate. 
In such a case, $\xi$-dependence disappears since $R_{ZZ}(\theta)$ commute with $r_Z(\pm\xi)$. 
On the contrary, when $\tilde{\Theta} \ne \Theta^*$, the deviations of $\tilde{\Theta}$ from the optimal parameter set $\Theta^*$ are 
reflected in the $\xi$-dependence. 
To amplify the $\xi$-dependence, we apply the operation $\tilde{id}(\theta, \xi)$ repeatedly $n$ times, denoted as $\tilde{id}(\theta, \xi)^n$.

The following is the procedure of parameter fitting using this $\xi$-dependence amplified experiment. 
First, we operate the gate set, $\{\tilde{id}(\theta, \xi_j)^n\}$ for ${\xi}_j = 2 \pi j/M\, (j=0,\cdots,M)$ with fixed $\theta$,   
onto $\ket{00}$ state and apply UTM to identify the set of unitary operations $\{U(\theta, \xi_j)\}$. 
By employing the parameterized unitary model, Eq.~(\ref{eq:U_Theta_t}), 
we define the cost function as
\begin{align}
    I^{(\theta)}(\Theta) = \frac{1}{M} \sum_{j=0}^M \tr[\{\tilde{id}(\theta, \xi_j)^n\}^\dagger \cdot U(\theta, \xi_j)], 
\end{align}
Note that the parameter set $\Theta$ is introduced in $\tilde{id}(\theta, \xi_j)$ by replacing $U_{CR}(d)$ in Eq.~(\ref{eq:Q_theta}) 
with the parameterized model unitary, Eq.~(\ref{eq:U_Theta_t}). 
We applied UTM to $\xi$-scan experiments for different $\theta$ values, $\{\theta_k\}\,{(k=1,2,\cdots,K)}$ and 
define the total cost function as 
\begin{align}
    I_{\rm total}(\Theta) = \sum_{k=1}^K I^{(\theta_k)}(\Theta).
\label{eq:total_cost_F}
\end{align}

\begin{figure}[ht]
\centering
\includegraphics[clip, width=8.5cm]{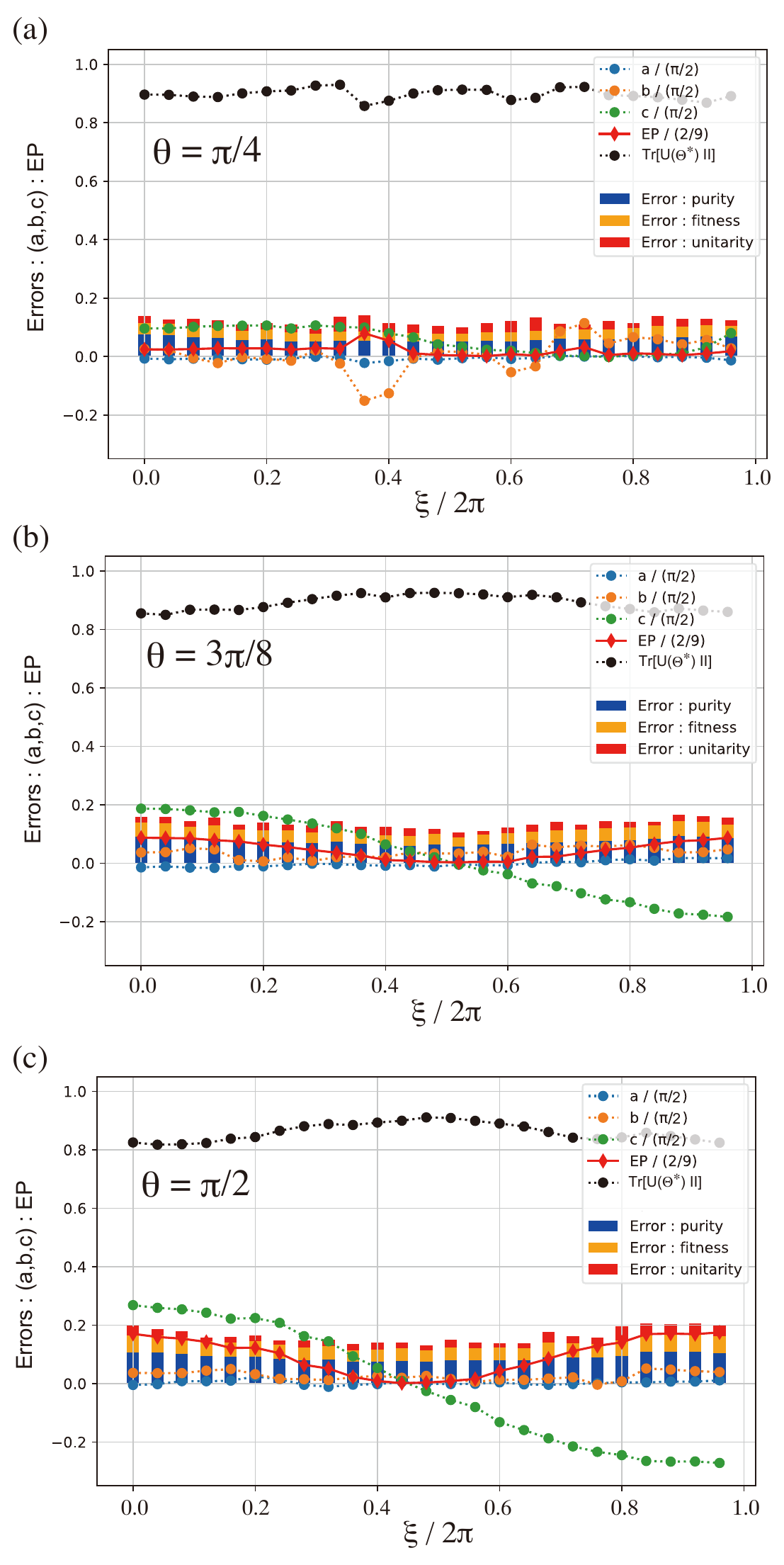}
\caption{UTM results of pseudo-identity gate $\tilde{id}(\theta, \xi)^n$ with $n=5$ for different $\theta$ values. 
(a) $\theta = \pi/4$, (b) $\theta = 3 \pi/8$, (c) $\theta = \pi/2$.
Blue, orange, and green dots with broken lines denote Cartan coefficients $a$, $b$, and $c$ values divided by $\pi/2$, respectively. 
The red line denotes the corresponding entangling power EP divided by $2/9$. 
Black dots with broken line denotes the trace fidelity between $U(\theta, \xi_j)$ and $I \otimes I$.
}
\label{fig:fine_UTM_experiments}
\end{figure}

Figure~\ref{fig:fine_UTM_experiments} presents the UTM results of $\tilde{id}(\theta, \xi_j)^n$ with $n=5$ for $\theta = \pi/4$, $3\pi/8$, and $\pi/2$.
We obtained the optimal values in Table~\ref{tab:U_H_parameters} by minimizing the total cost function Eq.~(\ref{eq:total_cost_F}) 
evaluated with the results in Fig.~\ref{fig:fine_UTM_experiments}. 
For all parameter fitting procedures, we utilized sequential least squares programming (SLSQP) to solve non-linear optimization problems.

\subsection{Device/Qiskit library information}\label{app:device_info}
\begin{table}[ht]
\caption{device information of {\tt ibm\_kawasaki} and the versions of qiskit libraries}
\label{tab:device_info}
 \centering
  \begin{tabular}{cll}
\\
   \hline
   &{\bf Device information}     &  \\
   \hline \hline
   &Device name                  & {\tt ibm\_kawasaki}\\
   &Number of qubits             & 27 \\
   &Quantum Volume               & 128\\
   &Processor type               & Falcon r5.11  \\
   &Basis gates                  & {\tt CX, ID, RZ, SX, X} \\
   &Median CX error              & $6.832\times10^{-3}$ \\
   &Median SX error              & $2.273\times10^{-4}$ \\
   &Median readout error         & $1.160\times10^{-2}$ \\
   &Median T1                    & $116.15 \mu$s \\
   &Median T2                    & $115.54 \mu$s \\
\\
   \hline
   &{\bf Qiskit libraries}              & version \\
   \hline \hline
   &{\tt qiskit-terra}           & 0.22.0 \\
   &{\tt qiskit-aer}             & 0.11.0 \\
   &{\tt qiskit-ignis}           & 0.7.1  \\
   &{\tt qiskit-ibmq-provider}   & 0.19.2 \\
   &{\tt qiskit}                 & 0.39.0 \\
   \hline
  \end{tabular}
\end{table}

\EOD
\end{document}